\documentclass[preprintnumbers,article,amsmath,amssymb,floatfix,10pt,prd,onecolumn,
superscriptaddress,nofootinbib]{revtex4-2}
\usepackage{bm}
\usepackage{amsfonts}
\usepackage{latexsym}
\usepackage[latin1]{inputenc}
\usepackage{graphicx}
\usepackage{amsmath}
\usepackage{palatino}
\usepackage{mathpazo}
\usepackage{textcomp}
\usepackage{float}
\usepackage{booktabs}
\usepackage{dcolumn}
\usepackage{ragged2e}
\usepackage{hyperref}
\hypersetup{colorlinks,citecolor=blue}
\hypersetup{colorlinks=true,linkcolor=red,filecolor=magenta,    urlcolor=blue}
\usepackage{amsmath}
\usepackage{xcolor}
\usepackage{orcidlink}
\usepackage{epsfig}
\usepackage{caption}
\usepackage{subcaption}
\usepackage{commath}
\def\jnl@style{\it}
\def\aaref@jnl#1{{\jnl@style#1}}

\def\aaref@jnl#1{{\jnl@style#1}}

\def\aj{\aaref@jnl{AJ}}                   
\def\apj{\aaref@jnl{ApJ}}                 
\def\apjl{\aaref@jnl{ApJ}}                
\def\apjs{\aaref@jnl{ApJS}}               
\def\apss{\aaref@jnl{Ap\&SS}}             
\def\aap{\aaref@jnl{A\&A}}                
\def\aapr{\aaref@jnl{A\&A~Rev.}}          
\def\aaps{\aaref@jnl{A\&AS}}              
\def\mnras{\aaref@jnl{Mon.~Not.~Roy.~Astron.~Soc.}}             
\def\prd{\aaref@jnl{Phys.~Rev.~D}}        
\def\prc{\aaref@jnl{Phys.~Rev.~C}}  
\def\prl{\aaref@jnl{Phys.~Rev.~Lett.}}    
\def\qjras{\aaref@jnl{QJRAS}}             
\def\skytel{\aaref@jnl{S\&T}}             
\def\ssr{\aaref@jnl{Space~Sci.~Rev.}}     
\def\zap{\aaref@jnl{ZAp}}                 
\def\nat{\aaref@jnl{Nature}}              
\def\aplett{\aaref@jnl{Astrophys.~Lett.}} 
\def\apspr{\aaref@jnl{Astrophys.~Space~Phys.~Res.}} 
\def\physrep{\aaref@jnl{Phys.~Rep.}}      
\def\physscr{\aaref@jnl{Phys.~Scr}}       
\def\commat{\aaref@jnl{Comm.~Math.~Phys.}}              
\def\science{\aaref@jnl{Science}}               
\def\cqg{\aaref@jnl{Classical Quant.~Grav.}}            
\def\jpcs{\aaref@jnl{JPCS}}                                     
\def\ijmpd{\aaref@jnl{Int.~J.~Mod.~Phys.~D}}                    
\def\grg{\aaref@jnl{Gen.~Relat.~Gravit.}}               
\def\rpp{\aaref@jnl{Rep.~Prog.~Phys.}}          
\def\npa{\aaref@jnl{Nucl.~Phys.~A}}        
\def\lrr{\aaref@jnl{Living Rev.~Rel.}}                   
\def\jcap{\aaref@jnl{J.~Cosmology Astropart.~Phys.}}    
\def\rmp{\aaref@jnl{Rev.~Mod.~Phys.}}   
\def\epjc{\aaref@jnl{Eur.~Phys.~J.~C}} 
\def\plb{\aaref@jnl{~Phy.~Lett.~B}} 
\def\mpla{\aaref@jnl{Mod.~Phy.~Lett.~A}} 
\def\arxiv{\aaref@jnl{arxiv.org}}

\allowdisplaybreaks[1]

\begin{document}
\color{black}       
\title{Investigating the compatibility of exact solutions in Weyl-type $f(Q,T)$ gravity with observational data}

\author{M. Koussour\orcidlink{0000-0002-4188-0572}}
\email[Email: ]{pr.mouhssine@gmail.com}
\affiliation{Department of Physics, University of Hassan II Casablanca, Morocco.}

\author{S. Myrzakulova\orcidlink{0000-0002-0027-0970}}
\email[Email: ]{shamyrzakulova@gmail.com}
\affiliation{L. N. Gumilyov Eurasian National University, Astana 010008,
Kazakhstan.}

\author{N. Myrzakulov\orcidlink{0000-0001-8691-9939}}
\email[Email: ]{nmyrzakulov@gmail.com}
\affiliation{L. N. Gumilyov Eurasian National University, Astana 010008,
Kazakhstan.}

%

\begin{abstract}
In this study, we investigate the dynamics of the Universe during the observed late-time acceleration phase within the framework of the Weyl-type $f(Q,T)$ theory. Specifically, we consider a well-motivated model with the functional form $f(Q,T)=\alpha Q+\frac{\beta }{6\kappa ^2}T$, where $Q$ represents the scalar of non-metricity and $T$ denotes the trace of the energy-momentum tensor. 
In this context, the non-metricity $Q_{\mu\alpha\beta}$ of the space-time is established by the vector field $w_\mu$. The parameters $\alpha$ and $\beta$ govern the gravitational field and its interaction with the matter content of the Universe. By considering the case of dust matter, we obtain exact solutions for the field equations and observe that the Hubble parameter $H(z)$ follows a power-law behavior with respect to redshift $z$. To constrain the model parameters, we analyze various datasets including the $Hubble$, $Pantheon$ datasets, and their combination. Our results indicate that the Weyl-type $f(Q,T)$ theory offers a viable alternative to explain the observed late-time acceleration of the Universe avoiding the use of dark energy. 

\end{abstract}
\date{\today}
\maketitle

\section{Introduction}
\label{sec1}

The finding of the accelerating expansion of the Universe in modern observational astrophysics presents a significant puzzle. According to our understanding of gravity as an attractive force, the presence of matter should decelerate the expansion velocity. However, observations of Type Ia Supernovae (SNe Ia) by Riess et al. \cite{Riess/1998,Riess/2004} and Perlmutter et al. \cite{Perlmutter/1999} suggest that the Universe is primarily composed of an exotic component known as Dark Energy (DE) responsible for the Universe's acceleration, accounting for approximately 70\% of its energy content. Astronomical observations from multiple sources, including CMBR \cite{Komatsu/2011,Huang/2006}, and LSS \cite{Koivisto/2006, Daniel/2008}, have yielded compelling evidence for a significant transition in the evolution of the Universe. These observations indicate a shift from an early deceleration phase to a more recent phase characterized by accelerated expansion. DE possesses an equation of state (EoS) with $\omega<-1/3$, justifying the apparent anti-gravitational nature of the current Universe dynamics \cite{Bambar}. In the framework of the standard $\Lambda$CDM cosmology, DE is mathematically represented by a cosmological constant in Einstein's field equations, with $\omega=-1$. However, the $\Lambda$CDM model faces various challenges, including the problems of cosmic coincidence, fine-tuning, dark matter, big bang singularity, and other limitations outlined in Clifton et al. \cite{Clifton} and related literature \cite{Peebles/2003,Sahni/2000,CdM1,CdM2}. These shortcomings have motivated the exploration of alternative cosmological models as potential solutions to these issues.

Therefore, it is necessary to explore alternative forms of matter that can explain the observed phenomena without the limitations of the $\Lambda$CDM model. One such possibility is the introduction of scalar fields that slowly roll down their potential, generating sufficient negative pressure to drive the accelerated expansion of the Universe. These alternative models, known as dynamical models of DE, provide an alternative explanation for the accelerated expansion of the Universe while addressing the shortcomings of the $\Lambda$CDM model.

An intriguing possibility to explain the observed phenomena is to consider the breakdown of the Einstein gravity model at large scales and introduce a more general framework to describe the gravitational field. This has led to the exploration of theoretical models where the standard Einstein-Hilbert (EH) action is replaced by a function of the Ricci scalar $R$ \cite{Nojiri2011}. Recently, there has been extensive investigation into these models, known as $f(R)$ gravity, which offer a promising explanation for the late-time cosmic acceleration of the Universe \cite{Carroll2004}. To understand the phenomenon of cosmic acceleration, numerous modified theories of gravity (MTGs) have emerged, including $f(R, T)$ gravity, $f(\mathcal{T})$ gravity, $f(G)$ gravity, $f(Q)$ gravity, and others. 

In the literature, General Relativity (GR) can be considered in three equivalent formulations. The first is the curvature formulation, where both torsion $\mathcal{T}$ and non-metricity $Q$ are assumed to be zero. This includes various MTGs such as $f(R)$ \cite{Nojiri2011,Carroll2004}, $f(R, T)$ \cite{Harko2011,Koussour1,B2}, and $f(G)$ gravity \cite{Felice2009,Koussour2}. In these theories, the gravitational field is characterized by a modified action that is a function of the Ricci scalar ($R$), the Ricci scalar coupled with the trace of the energy-momentum tensor ($R$, $T$), or the Gauss-Bonnet scalar ($G$), respectively. The second formulation of GR is the teleparallel representation, which is solely based on the concept of torsion. In this framework, the gravitational field is characterized by a MTG known as $f(T)$ theory \cite{Hayashi,bengocheu09,linder10, Koussour3,Benetti}. Lastly, there is the symmetric teleparallel formulation, where gravity is connected to non-metricity, as seen in theories like $f(Q)$ theory \cite{fQ1,fQ2,fQ4,fQ5,fQ6,fQ7,fQ8}. In an effort to combine electromagnetism and gravity \cite{fQ3}, Weyl proposed a Riemannian extension to GR. In this extension, the non-metricity of space-time gives rise to the electromagnetic field. When vectors are parallel transported, their orientation and length can vary. The non-metricity of space-time precisely describes this variation in the length of a vector. An alternative and recent extension of $f(Q)$ theory is $f(Q, T)$ theory \cite{fQT1}. This modified theory incorporates a non-minimal coupling in the gravitational action, where the conventional Lagrangian is replaced by an arbitrary function, that depends on both the non-metricity $Q$ and the trace of the energy-momentum tensor $T$. The inclusion of $T$ dependence in these models can be attributed to various factors, including exotic imperfect fluids and certain quantum effects. In $f(Q, T)$ theory, there is a significant advancement over GR through the coupling of matter and geometry. This coupling leads to a non-zero covariant divergence of the energy-momentum tensor, resulting in the motion of test particles deviating from geodesic paths. These models have been successful in explaining the late-time cosmic accelerated expansion of the Universe \cite{QT2,QT3}. By varying the gravitational field equations action in relation to the metric tensor, one can derive the field equations of the $f(Q, T)$ theory. Several authors, including Najera and Fajardo \cite{QT4}, Shiravand et al. \cite{QT5}, and Bourakadi et al. \cite{QT6}, among others, have extensively studied several cosmological concepts within the framework of $f(Q, T)$ theory. Arora et al. \cite{QT7} derived exact solutions for the FLRW cosmological model, which is filled with a perfect fluid matter, within the framework of $f(Q, T)$ theory. T. H. Loo \cite{QT8} investigated the properties and evolution of the Bianchi type-I cosmological model, which was filled with a perfect fluid. On the other hand, Tayde et al. \cite{QT9} focused on the modeling of static wormholes within the framework of $f(Q, T)$ theory.  

In this study, we examine the exact solutions of a specific formulation of the $f(Q, T)$ gravity that focuses on the non-minimal coupling between $Q$ and $T$. Our investigation is carried out within the framework of proper Weyl geometry, where we adopt a particular expression for the non-metricity $Q$ derived from the non-conservation of the metric tensor's divergence, $\nabla _{\mu }g_{\alpha \beta
}=-w_{\mu }g_{\alpha \beta }$. By employing this approach, we describe the non-metricity in terms of a vector field $w_{\mu }$, which, in conjunction with the metric tensor, fully determines the non-metricity. This formulation is commonly referred to as Weyl-type $f(Q, T)$ gravity \cite{Weyl1}. Our goal is to explore the constraints and implications of the Weyl-type $f(Q, T)$ gravity, with its specific coupling between the non-metricity and the trace of the matter energy-momentum tensor. Previous studies by Yang et al. \cite{Weyl2} extensively investigated various aspects of Weyl-type $f(Q, T)$ gravity. They focused on geodesic deviation, the Raychaudhuri equation, the Newtonian limit, and tidal forces within this framework. Further, Koussour \cite{Weyl3} introduced a model-independent approach in Weyl-type $f(Q, T)$ gravity to study the crossing of the phantom divide line. The author examined the behavior of the crossing of the phantom divide of the function $f(Q, T)=\alpha Q+\frac{\beta }{6\kappa ^2}T$. The phantom divide line refers to the transition between different cosmic expansion phases, specifically from decelerated expansion to accelerated expansion with a phantom-like EoS. Our goal in this study is to present the full constraints of the Weyl-type $f(Q, T)$ model. To achieve this, we employ a comprehensive set of observational data, including $Hubble$ and $Pantheon$ datasets.

This study is structured as follows: Sec. \ref{sec2} provides an overview of Weyl-type $f(Q,T)$ gravity. In Sec. \ref{sec3}, we investigate the exact solutions of the modified Friedmann equations in Weyl-type $f(Q,T)$ gravity. Specifically, we consider the well-motivated form of the function $f(Q, T)$ given by $f(Q,T)=\alpha Q+\frac{\beta }{6\kappa ^2}T$, where $\alpha$ and $\beta$ are model parameters. We focus on the case of dust matter and determine the Hubble parameter $H(z)$ as a function of redshift $z$. In Sec. \ref{sec4}, we use a dataset consisting of 31 $Hubble$ data points and 1048 $Pantheon$ data points to constrain the unknown parameters $H_0$, $\alpha$, $\beta$, and $M$. We perform a comparison between the Weyl-type $f(Q,T)$ model and the $\Lambda$CDM model using error bar plots. Further, we investigate the evolution of the density parameter and deceleration parameter within our framework. Finally, Sec. \ref{sec5} presents the results obtained and provides a comprehensive discussion.

\section{Overview of Weyl-type $f(Q,T)$ gravity}
\label{sec2}

The action in Weyl-type $f(Q,T)$ gravity is given as \cite{Weyl1,Weyl2} 
\begin{equation}
S=\int \sqrt{-g}d^{4}x \left[ \kappa ^{2}f(Q,T)-\frac{1}{4}W_{\alpha \beta }W^{\alpha \beta
}-\frac{1}{2}m^{2}w_{\alpha }w^{\alpha }+\lambda (R+6\nabla _{\mu }w^{\mu }-6w_{\mu }w^{\mu })+\mathcal{L}_{m}%
\right].
\label{1}
\end{equation}

Here, the field strength tensor of the vector field is represented as $W_{\alpha \beta }=\nabla _{\beta }w_{\alpha }-\nabla _{\alpha
}w_{\beta \text{ }}$, where $w_{\alpha}$ denotes the vector field itself. The value of $\kappa ^{2}$ is defined as $1/16\pi G$, where $G$ represents the gravitational constant, and $m$ signifies the particle mass associated with the vector field. The matter Lagrangian is denoted as $\mathcal{L}_{m}$.

In addition, the action mentioned above consists of three terms. The first term corresponds to the gravitational interaction described by the Weyl-type $f(Q,T)$ function. The second term represents the ordinary kinetic term of the vector field, while the third term accounts for the mass term associated with the vector field. It is important to note that the function $f(Q,T)$ represents an arbitrary function of the non-metricity scalar $Q$ and the trace of the matter-energy-momentum tensor $T$. Further, the Lagrange multiplier $\lambda$ is introduced as a parameter to enforce constraints or conditions within the theory, particularly the imposition of the flat geometry constraint, where the total curvature of the Weyl space vanishes i.e. $\bar{R}=0$.

The non-metricity scalar is defined as, 
\begin{equation}
Q\equiv -g^{\alpha \beta }\left( L_{\nu \beta }^{\mu }L_{\beta \mu }^{\nu
}-L_{\nu \mu }^{\mu }L_{\alpha \beta }^{\nu }\right) ,  \label{2}
\end{equation}%
where $L_{\alpha \beta }^{\lambda }$ represents the tensor of deformation and is expressed as
\begin{equation}
L_{\alpha \beta }^{\lambda }=-\frac{1}{2}g^{\lambda \gamma }\left( Q_{\alpha
\gamma \beta }+Q_{\beta \gamma \alpha }-Q_{\gamma \alpha \beta }\right) .
\label{3}
\end{equation}

In Riemannian geometry, the Levi-Civita connection $\Gamma _{\alpha
\beta }^{\lambda }$ and the metric tensor $g_{\alpha\beta}$ can both be compatible, meaning that their covariant derivative with respect to the connection vanishes, i.e., $\nabla_{\mu} g_{\alpha\beta} = 0$. However, in Weyl's geometry, this compatibility appears to be modified. The presence of non-metricity captures the departure from metric compatibility, offering insights into the geometric properties of the space-time under investigation. Therefore, we have the following relation:
\begin{equation}
\overline{Q}_{\mu \alpha \beta }\equiv \overline{\nabla }_{\mu }g_{\alpha
\beta }=\partial _{\mu }g_{\alpha \beta }-\overline{\Gamma }_{\mu \alpha
}^{\rho }g_{\rho \beta }-\overline{\Gamma }_{\mu \beta }^{\rho }g_{\rho
\alpha }=2w_{\mu }g_{\alpha \beta },  \label{4}
\end{equation}%
where, 
\begin{equation}
    \overline{\Gamma }_{\alpha \beta }^{\lambda }\equiv \Gamma _{\alpha
\beta }^{\lambda }+g_{\alpha \beta }w^{\lambda }-\delta _{\alpha }^{\lambda
}w_{\beta }-\delta _{\beta }^{\lambda }w_{\alpha },
\end{equation}
and $\Gamma _{\alpha
\beta }^{\lambda }$ represent the semi-metric connection in Weyl geometry and the Christoffel symbol in terms of the metric
tensor $g_{\alpha \beta }$, respectively. Weyl introduced the semi-metric connection to capture the joint variation of both direction and magnitude experienced by a vector field.

By using Eqs. \eqref{2}-\eqref{4}, we can derive the following relationship:
\begin{equation}
Q=-6w^{2}.  \label{5}
\end{equation}

Now, by taking the variation of the action with respect to the vector field, we can derive the generalized Proca equation,
\begin{equation}\label{EOM1}
    \nabla^\beta W_{\alpha \beta }-(m^2+12\kappa^2 f_Q+12\lambda)w_\alpha=6\nabla_\alpha \lambda.
\end{equation}

By comparing Eq. (\ref{EOM1}) with the standard Proca equation, it becomes evident that the effective dynamical mass of the vector field can be expressed in as
\begin{equation}\label{discussion1}
m^2_{\rm{eff}}=m^2+12\kappa^2f_Q+12\lambda.
\end{equation}
where 
\begin{equation}
    f_{Q}\equiv \frac{%
\partial f(Q,T)}{\partial Q}. 
\end{equation}

It is important to note that the Lagrange multiplier field gives rise to an effective current for the vector field. In the realm of quantum field theory, experimental measurements often reveal discrepancies between the observed mass and the bare mass, which can be attributed to the presence of interactions. Therefore, Eq. (\ref{discussion1}) emphasizes that within the framework of Weyl-type $f(Q,T)$ gravity, these deviations in mass can also arise from the nontrivial geometric properties of the space-time.

Furthermore, the generalized field equation is obtained by performing the variation of the action \eqref{1} with respect to the metric tensor,
\begin{multline}
\frac{1}{2}\left( T_{\alpha \beta }+S_{\alpha \beta }\right) -\kappa
^{2}f_{T}\left( T_{\alpha \beta }+\Theta _{\alpha \beta }\right) =-\frac{%
\kappa ^{2}}{2}g_{\alpha \beta }f
-6\kappa ^{2}f_{Q}w_{\alpha }w_{\beta }+\lambda \left( R_{\alpha \beta
}-6w_{\alpha }w_{\beta }+3g_{\alpha \beta }\nabla _{\rho }w^{\rho }\right) 
\\
+3g_{\alpha \beta }w^{\rho }\nabla _{\rho }\lambda -6w_{(\alpha }\nabla
_{\beta )}\lambda +g_{\alpha \beta }\square \lambda -\nabla _{\alpha }\nabla
_{\beta }\lambda ,
\label{7}
\end{multline}%
where 
\begin{equation}
T_{\alpha \beta }\equiv -\frac{2}{\sqrt{-g}}\frac{\delta (\sqrt{-g}L_{m})}{%
\delta g^{\alpha \beta }},  \label{8}
\end{equation}%
and%
\begin{equation}
f_{T}\equiv \frac{\partial f(Q,T)}{\partial T},  \label{9}
\end{equation}%
respectively. Furthermore, we introduce the quantity $\Theta_{\alpha\beta}$, which is defined as
\begin{equation}
\Theta _{\alpha \beta }\equiv g^{\mu \nu }\frac{\delta T_{\mu \nu }}{\delta
g_{\alpha \beta }}=g_{\alpha \beta }L_{m}-2T_{\alpha \beta }-2g^{\mu \nu }%
\frac{\delta ^{2}L_{m}}{\delta g^{\alpha \beta }\delta g^{\mu \nu }}.
\label{10}
\end{equation}

In the given field equation, $S_{\alpha\beta}$ represents the rescaled energy-momentum tensor associated with the free Proca field.
\begin{equation}
S_{\alpha \beta }=-\frac{1}{4}g_{\alpha \beta }W_{\rho \sigma }W^{\rho
\sigma }+W_{\alpha \rho }W_{\beta }^{\rho }-\frac{1}{2}m^{2}g_{\alpha \beta
}w_{\rho }w^{\rho }+m^{2}w_{\alpha }w_{\beta }.  \label{11}
\end{equation}

In the context of the Weyl-type $f(Q,T)$ theory, the divergence of the matter-energy-momentum tensor can be expressed as \cite{Weyl1}
\begin{align}\label{div}
    \nabla^\alpha T_{\alpha\beta}=\frac{\kappa^2}{1+2\kappa^2 f_T}\Big [ 2\nabla_\beta(\mathcal{L}_{m}f_T)-f_T\nabla_\beta T-2T_{\alpha\beta}\nabla^\alpha f_T\Big ].
\end{align}

Hence, the equation presented above illustrates that within the framework of the Weyl-type $f(Q,T)$ theory, the matter energy-momentum tensor does not exhibit conservation. It is noteworthy to emphasize that when $f_T = 0$, the energy-momentum tensor becomes conserved.

\section{Cosmological Model and Exact Solutions in Weyl-type $f(Q,T)$ Gravity}
\label{sec3}

In this study, we investigate a Friedmann-Lemaitre-Robertson-Walker (FLRW) Universe, described by the isotropic, homogeneous, and spatially flat metric, 
\begin{equation}
ds^2 = -dt^2 + a^2(t) \left(dr^2 + r^2 d\Omega^2 \right).  \label{FLRW}
\end{equation}%

Here, $t$ denotes the cosmic time, $a(t)$ represents the scale factor that describes the expansion of the Universe, $r$ is the comoving radial coordinate, and $d\Omega^2$ represents the line element of the unit 2-sphere. Furthermore, we assume that the vector field can be parametrized as
\begin{equation}
w_{\alpha }=\left[
\psi (t),0,0,0\right].
\end{equation}

Hence, $w^{2}=w_{\alpha }w^{\alpha }=-\psi
^{2}(t)$ and $Q=-6w^{2}=6\psi ^{2}(t)$. Again, we consider the Universe to be described by a perfect fluid, for which the energy-momentum tensor is defined as
\begin{equation}
\label{EMT}
T_{\mu\nu}=\left(\rho+p\right)u_\mu u_\nu+ p g_{\mu\nu}.
\end{equation}

Here, $\rho$ represents the energy density, $p$ denotes the pressure, and $u^\alpha$ corresponds to the 4-velocity of the fluid, satisfying the condition $u_\alpha u^\alpha=-1$. This implies that $T^\alpha_\beta=diag\left(-\rho,p,p,p\right)$, and $\Theta^\alpha_\beta=\delta^\alpha_\beta p-2T^\alpha_\beta=diag\left(2\rho+p,-p,-p,-p\right)$.

In the context of cosmology, the requirements of a flat space-time and the generalized Proca equation can be expressed as
\begin{eqnarray}
\dot{\psi} &=&\dot{H}+2H^{2}+\psi ^{2}-3H\psi ,  \label{17} \\
\dot{\lambda} &=&\left( -\frac{1}{6}m^{2}-2\kappa ^{2}f_{Q}-2\lambda \right)
\psi =-\frac{1}{6}m_{eff}^{2}\psi ,  \label{18} \\
\partial _{i}\lambda &=&0.  \label{19}
\end{eqnarray}

Here, the Hubble parameter, denoted as $H(t) = \frac{\dot{a}}{a}$, signifies the rate at which the Universe is expanding. The dot (.) symbolizes the derivative with respect to time, denoted as $t$.

By using Eq. (\ref{7}) and incorporating the given metric (\ref{FLRW}), we obtain the generalized Friedmann equations as \cite{Weyl1}, 
\begin{equation}
\kappa ^{2}f_{T}(\rho +p)+\frac{1}{2}\rho =\frac{\kappa ^{2}}{2}f-\left(
6\kappa ^{2}f_{Q}+\frac{1}{4}m^{2}\right) \psi ^{2}-3\lambda (\psi ^{2}-H^{2})-3\dot{\lambda}(\psi -H),  \label{F1}
\end{equation}%
\begin{equation}
-\frac{1}{2}p =\frac{\kappa ^{2}}{2}f+\frac{m^{2}\psi ^{2}}{4}+\lambda
(3\psi ^{2}+3H^{2}+2\dot{H}) +(3\psi +2H)\dot{\lambda}+\ddot{\lambda}.  \label{F2}
\end{equation}

In the special scenario where $f=0$, $\psi=0$, and $\lambda=\kappa^2$, the gravitational action (\ref{1}) simplifies to the standard Hilbert-Einstein action. Consequently, the generalized equations (\ref{F1}) and (\ref{F2}) reduce to the Friedmann equations in GR. Specifically, these equations take the form, 
\begin{eqnarray}
   3H^2&=&\frac{\rho}{2\kappa^2}\\
    2\dot{H}&=&-\frac{(\rho+p)}{2\kappa^2}
\end{eqnarray}
respectively. In this limit, the energy density $\rho$ and pressure $p$ adhere to the standard framework of GR.

To investigate the exact solutions of Weyl-type $f(Q, T)$ gravity theory, we need to specify a functional form for $f(Q, T)$. In our analysis, we will focus on a specific functional form given by \cite{Weyl1, Weyl3}
\begin{equation}
    f(Q,T)=\alpha Q+\frac{\beta }{6\kappa ^2}T
    \label{fQT}
\end{equation}
where $\alpha$ and $\beta$ play a crucial role in determining the behavior and characteristics of the model. These parameters allow us to fine-tune the specific features and properties of the Weyl-type $f(Q, T)$ gravity theory. Furthermore, we consider the dimensionless quantity (Weyl field mass) $M^2= m^{2}/\kappa^2$, which characterizes the strength of the coupling between the Weyl geometry and matter. It is important to note that when $\beta=0$, the function $f(Q,T)$ reduces to $f(Q,T)=\alpha Q$, which corresponds to the well-established theory of GR. Further, when $T = 0$, representing the vacuum case, the theory simplifies to $f(Q)$ gravity, which is equivalent to GR and has been extensively tested and validated within the Solar System \cite{fQ1,fQ2}. Moreover, previous studies by Yixin et al. \cite{Weyl1,fQT1} have shown that the considered model exhibits an accelerating expansion of the Universe, ultimately leading to a de Sitter-like evolution. These findings provide further support for the viability and consistency of the proposed framework. Hence, $f_Q=\alpha$ and $f_T=\frac{\beta }{6\kappa ^2}$, respectively. When considering this particular form of the coupling function, the gravitational field equations, as expressed in Eqs. (\ref{F1})-(\ref{F2}), can be written as
\begin{equation}
\label{F11}
\frac{\beta}{12}p-\left(\frac{\beta }{4}+\frac{1}{2}\right) \rho =3 \alpha  \kappa ^2 \psi ^2+3 \kappa ^2 \left(\psi ^2-H^2\right)+\frac{m^2 \psi ^2}{4},
\end{equation}
\begin{equation}
\label{F22}
\frac{\beta }{12}\rho-\left(\frac{\beta }{4}+\frac{1}{2}\right) p=3 \alpha  \kappa ^2 \psi ^2+\kappa ^2 \left(3 H^2+2 \dot{H}+3 \psi ^2\right)+\frac{m^2 \psi ^2}{4}.
\end{equation}

To account for generality, we consider the cosmological matter to follow an equation of state given by $p = (\gamma-1)\rho$, where gamma is a constant with a value between 1 and 2. By solving Eqs. (\ref{F11}) and (\ref{F22}), we find the expression for the energy density $\rho$ as
\begin{equation}
    \rho=\frac{3 \psi ^2 \left(12 (\alpha +1) \kappa ^2+m^2\right)-36 H^2 \kappa ^2}{\beta  (\gamma -4)-6}.
\end{equation}

Here, for the sake of simplicity, we adopt the convenient choice of setting $\lambda = \kappa^2 = 1$. \cite{Weyl1}. By solving the aforementioned equation with the assumption $H(t) = \psi(t)$ and $\gamma = 1$ (dust matter), we obtain the expression for the energy density in terms of the Weyl field mass and Hubble parameter as
\begin{equation}
    \rho=-\frac{H^2 \left(12 \alpha +M^2\right)}{(\beta +2)}.
\end{equation}

Now, the dynamical equations (\ref{F11})-(\ref{F22}) that describe the evolution of the Universe can be expressed as
\begin{equation}
    \dot H+\chi H^2=0,
    \label{eqH}
\end{equation}
where 
\begin{equation}
    \chi=\frac{3 (\alpha +\beta +2)}{\beta +2}+\frac{2 \alpha  \beta }{\beta +2}+\frac{(2 \beta +3) M^2}{12 (\beta +2)}.
\end{equation}

In addition, to better align the theoretical results with cosmological observations, we introduce a new independent variable, the redshift $z$, instead of the usual time variable $t$. The cosmological redshift is defined in terms of the following relation:
\begin{equation}
    1+z=\frac{1}{a(t)}.
\end{equation}

By imposing the condition that the present-day value of the scale factor is one ($a(0) = 1$), we can normalize the scale factor. As a result, we can replace the derivatives with respect to time with derivatives with respect to cosmological redshift by using the following relation:
\begin{equation}
    \dot{H}=\frac{dH}{dt}=-\left(1+z\right)H(z)\frac{dH}{dz}.
\end{equation}

By solving Eq. (\ref{eqH}), we obtain the exact solution for $H(z)$, which can be expressed as
\begin{equation}
    H(z)=H_{0} (1+z)^{\chi},
\end{equation}
where $H_0=H(z=0)$ represents the present value of the Hubble parameter. Furthermore, we have discovered that the power-law expansion serves as a solution to the field equations in Weyl-type $f(Q,T)$ gravity. Power-law cosmology has been extensively studied in various modified theories of gravity, including $f(R,T)$ \cite{B2} and $f(Q,T)$ gravity \cite{QT7}, as it offers a compelling resolution to several outstanding issues \cite{QT8}.

Now, we proceed to constrain the model parameters ($\alpha$, $\beta$, $M$) as well as the Hubble constant $H_0$ using a range of observational measurements. This analysis allows us to construct a physically realistic cosmological model that is consistent with astrophysical observations. By comparing our model predictions with observational data, we can determine the values of these parameters that best fit the observations and provide a robust description of the Universe's evolution.

\section{Observational constraints and validation of the model predictions}
\label{sec4}

In this section, we proceed to analyze and validate the results obtained from the exact solutions of FLRW in Weyl-type $f(Q,T)$ gravity. The accurate determination of model parameters is crucial for understanding the cosmological implications. To achieve this, we employ observational data and statistical techniques to constrain the values of parameters such as $H_0$, $\alpha$, $\beta$, and $M$. In particular, we use the Markov Chain Monte Carlo (MCMC) method combined with the standard Bayesian approach \cite{Mackey/2013}. By applying the chi-squared function $\Tilde{\chi}^2$, we construct the probability function ($\mathcal{L}\propto e^{-\frac{\Tilde{\chi}^2}{2}}$) to assess the goodness of fit between the theoretical predictions and observational data. The MCMC technique allows us to explore the parameter space and derive the posterior distribution of the model parameters. Through this analysis, we aim to identify the parameter values that best fit the observed data and provide a reliable cosmological model consistent with astrophysical observations.

In addition, we focus on analyzing two datasets, namely $Hubble$ and $Pantheon$ data, in order to validate our model. To ensure a comprehensive exploration of parameter space, we adopt appropriate priors for the model parameters. Specifically, we consider the following priors: $60 < H_0 < 80$, which encompasses a range of possible values for the Hubble constant and ensures compatibility with observational datasets; $-10 < \alpha < +10$ and $-10 < \beta < +10$ for the free model parameters, allowing for a wide range of values; and $0 < M < 10$ to enforce the positivity of the mass of the Weyl field.

\subsection{$Hubble$ $dataset$}
The analysis of the Hubble parameter plays a crucial role in understanding the expansion of the Universe. One of the convenient ways to study the expanding Universe is by expressing it in terms of the cosmological redshift, which provides valuable insights in various cosmological contexts. In this study, we use the technique of determining the Hubble parameter from the differential age method, which has proven to be successful. Specifically, we make use of 31 data points of $Hubble$ obtained from this method in the redshift range $0.07 \leq z \leq 2.41$ \cite{Yu/2018,Moresco/2015,Sharov/2018}. These data points serve as valuable constraints for our model and allow us to estimate the unknown parameters. To evaluate the unknown parameters, we employ the $\chi^2$ square function, which is commonly used in statistical analyses. In the case of the $Hubble$ dataset, the chi-square function takes the following form:
\begin{equation}
\chi^2_{Hubble}(H_0,\alpha,\beta,M)=\sum\limits_{k=1}^{31}\dfrac{\left[H_{th}(z_k,H_0,\alpha,\beta,M)-H_{ob}(z_k)\right]^2}{\sigma^2_{H(z_k)}}.
\end{equation}

Here, $H_{ob}$ represents the observed Hubble parameter at a particular redshift $z_k$, $H_{th}$ is the corresponding theoretical value predicted by our model, and $\sigma$ represents the uncertainty associated with the observed data.

\subsection{$Pantheon$ $dataset$}
The study of SNe Ia plays a crucial role in our understanding of the expanding Universe. These SNe, which can be spectroscopically observed, provide valuable data for studying cosmic expansion. Various surveys and observations, such as the SuperNova Legacy Survey (SNLS), Sloan Digital Sky Survey (SDSS), Hubble Space Telescope (HST) survey, and Panoramic Survey Telescope and Rapid Response System (Pan-STARRS1), have collected extensive data on SNe Ia, offering solid evidence for the expansion of the Universe. In this study, we use the $Pantheon$ dataset, which consists of 1048 magnitude measurements of the distance modulus of SNe Ia. These measurements were obtained over a redshift range of $0.01 \leq z \leq 2.3$ \cite{Scolnic/2018,Chang/2019}. The distance modulus, denoted by $\mu(z_k)$, is a theoretical quantity that describes the apparent brightness of the supernovae at different redshifts.

To validate our model and find the best fits, we compare the theoretical predictions of the distance modulus $\mu(z_k)$ with the observed values from the $Pantheon$ dataset. This involves performing a systematic comparison between the theoretical and observational data points. So, we have
 \begin{equation}
\mu^{th}(z_k)=\mu_0+5\,\text{log}_{10}(\mathcal{D}_L(z_k)),
\end{equation}
with the nuisance parameter is expressed as
\begin{equation}
\mu_0= 25+5\, \text{log}_{10}\left(\dfrac{1}{H_0 Mpc}\right),
\end{equation}
and the luminosity distance
\begin{equation}
\mathcal{D}_L(H_0,\alpha,\beta,M,z)=(1+z)\int\limits_0^z \dfrac{c}{H(H_0,\alpha,\beta,M,x)}dx.
\end{equation}

The expression for the $\Tilde{\chi}^2$ function corresponding to the Pantheon data sample, considering the covariance matrix $\mathcal{C}_{Pantheon}$, can be written as
\begin{equation}
\Tilde{\chi}^2_{Pantheon}(\mu_0,H_0,\alpha,\beta,M)=\sum\limits_{k,l=1}^{1048}\bar{\mu}_k \left(\mathcal{C}_{Pantheon}^{-1}\right)_{kl}\bar{\mu}_l,
\end{equation}
where, $\bar{\mu}_k=\mu^{th}(z_k,H_0,\alpha,\beta,M)-\mu^{ob}(z_k)$.

\subsection{$Hubble+Pantheon$ $dataset$}
In the previous subsections, we analyzed two distinct datasets, namely $Hubble$ and $Pantheon$. In this subsection, we will combine these two sets of sample points to investigate the best-fit for our model using the $Hubble+Pantheon$ dataset. To achieve this, we use the $\Tilde{\chi}^2$ function, which is expressed as
\begin{equation}
\Tilde{\chi}^2_{\text{Joint}}=\Tilde{\chi}^2_{Hubble}+\Tilde{\chi}^2_{Pantheon}.
\end{equation}

\begin{figure}[h]
\centering
\includegraphics[scale=0.45]{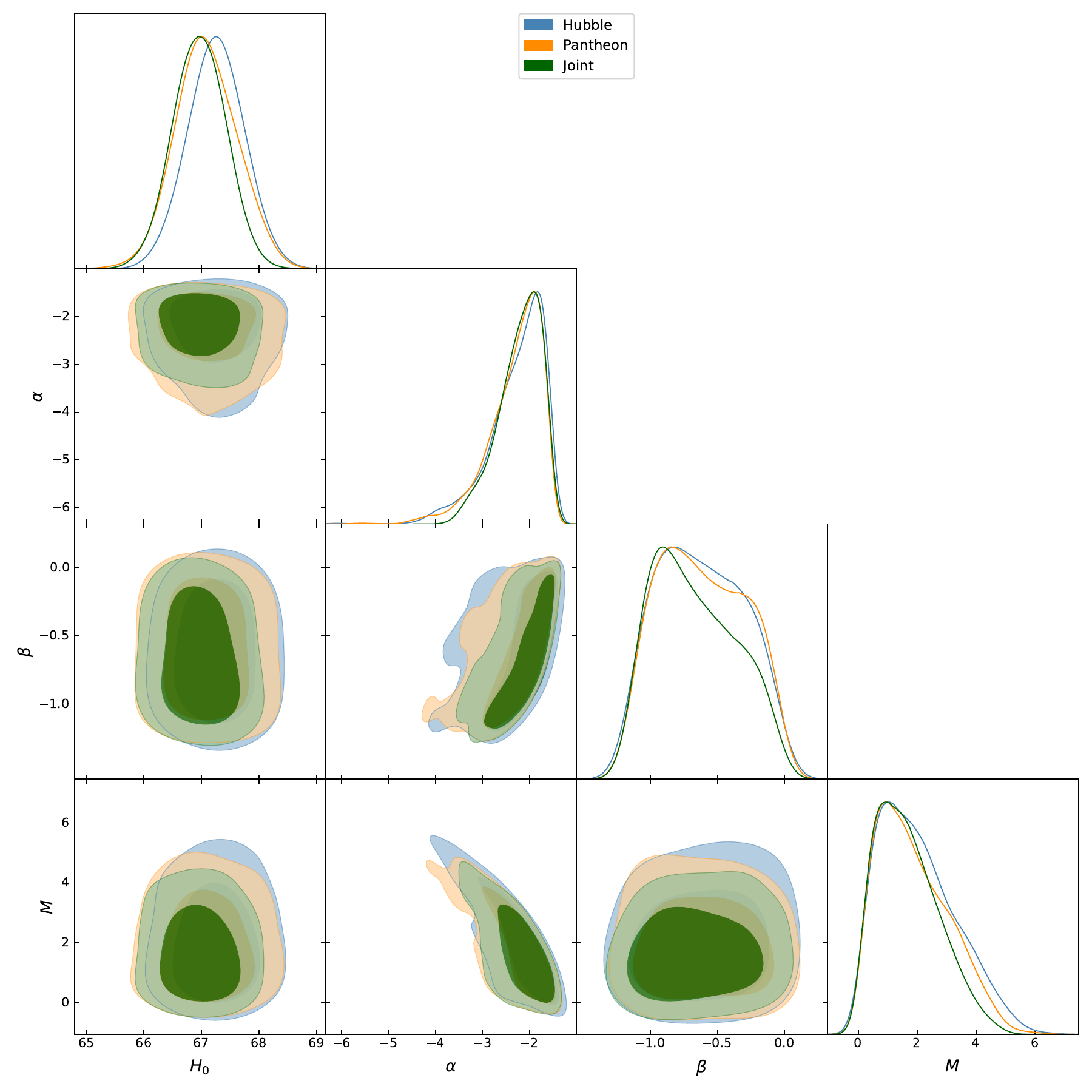}
\caption{The figure shows the contour plot of the unknown parameters $H_0$, $\alpha$, $\beta$, and $M$ with $1-\sigma$ and $2-\sigma$ errors, constraints from $Hubble$, $Pantheon$ and $Hubble+Pantheon$ datasets.}
\label{Combine}
\end{figure}

\begin{figure}[h]
\centering
\includegraphics[scale=0.60]{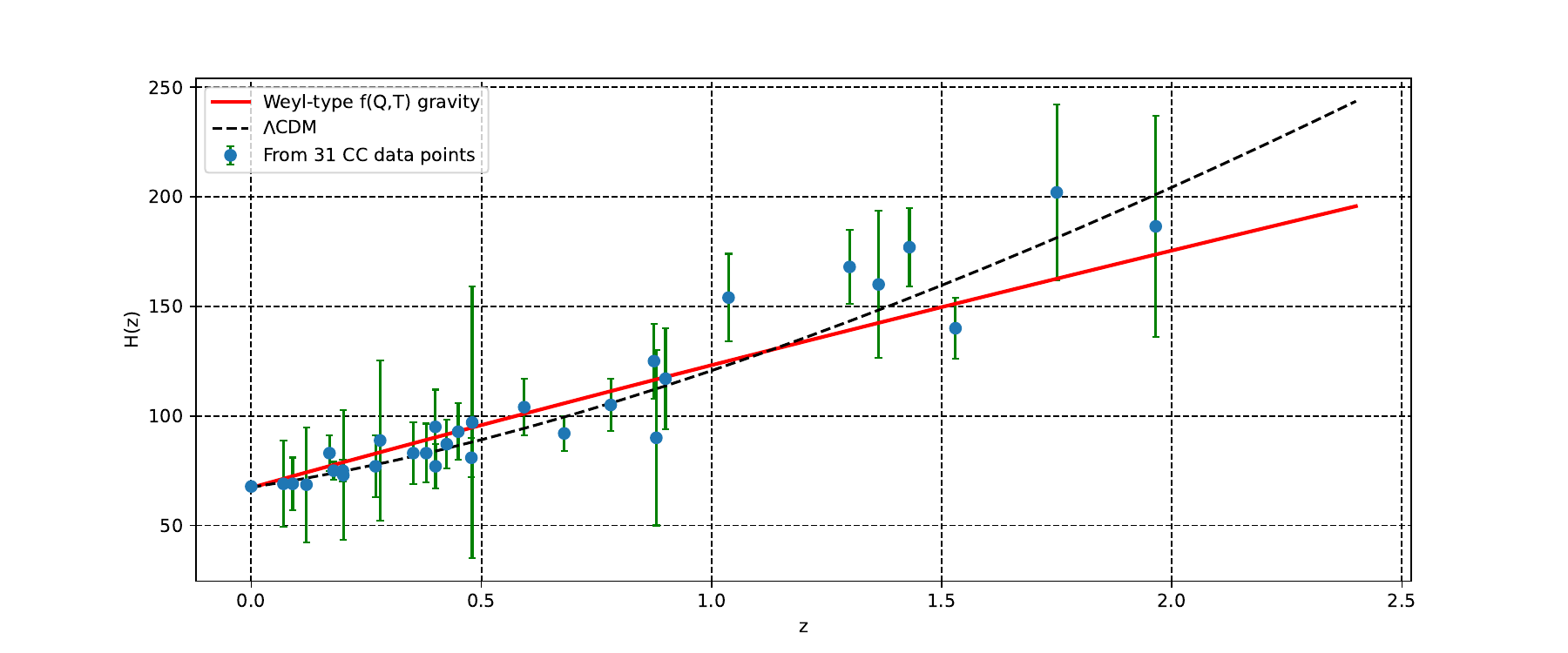}
\caption{The figure shows a comparison of the Hubble function $H(z)$ as a function of redshift $z$. Our model is represented by the green line, fitted to the 31 data points of the $Hubble$ dataset shown as dots with error bars. The $\Lambda$CDM model is depicted by the black solid line.}
\label{Hubble}
\end{figure}

\begin{figure}[h]
\centering
\includegraphics[scale=0.60]{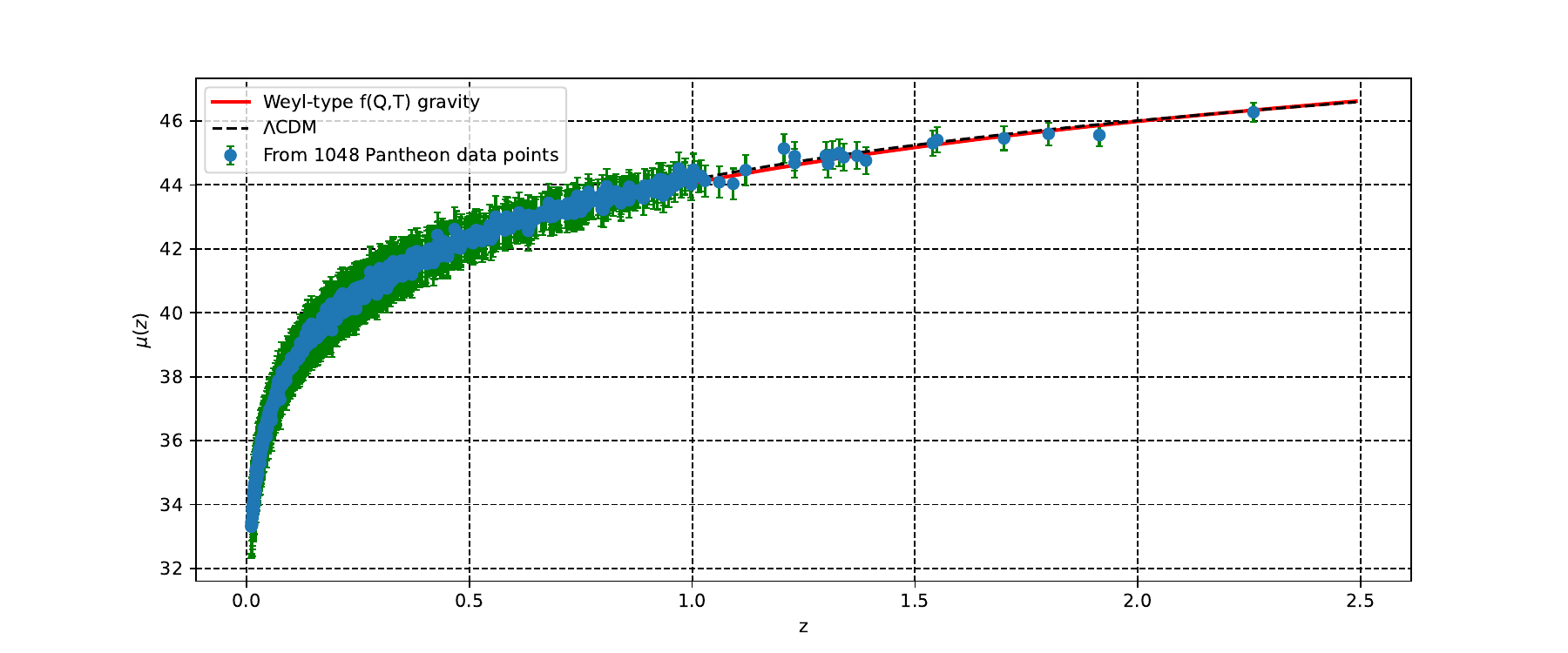}
\caption{The figure shows a comparison of the distance modulus $\mu(z)$ as a function of redshift $z$. Our model is represented by the green line, fitted to the 1048 data points of the $Pantheon$ dataset shown as dots with error bars. The $\Lambda$CDM model is depicted by the black solid line.}
\label{Mu}
\end{figure}

The analysis of the model parameters is performed by minimizing their respective $\Tilde{\chi}^2$ values using the MCMC method implemented with the emcee library. The resulting parameter constraints are summarized in Tab. \ref{tab}. Also, contour plots in Fig. \ref{Combine} illustrate the 1-$\sigma$ and 2-$\sigma$ confidence regions, offering a comprehensive visualization of the joint analysis of the $Hubble$, $Pantheon$, and joint observational datasets. Figs. \ref{Hubble} and \ref{Mu} showcase the fits of our proposed model alongside the $\Lambda$CDM model, which assumes $\Omega_{m0} = 0.315$, and $H_0 = 67.4$ km/s/Mpc. The error bars in these figures represent the uncertainties in the data points. These results provide valuable insights into the consistency between the model predictions and the observed data. In the next section, we will discuss in more detail these results and compare them with previous studies.

\begin{table*}[!htbp]
\begin{center}
\begin{tabular}{l c c c c c}
\hline\hline 
$datasets$              & $H_{0}$ ($km/s/Mpc$) & $\alpha$ & $\beta$ & $M$ & $q_{0}$ \\
\hline

$Hubble$ & $67.25_{-0.98}^{+0.97}$  & $-2.26^{+0.90}_{-1.3}$  & $-0.62^{+0.59}_{-0.56}$ & $2.0^{+2.6}_{-2.1}$ & $-0.46^{+2.14}_{-2.19}$ \\

$Pantheon$   & $67.1_{-1.0}^{+1.1}$  & $-2.29_{-1.2}^{+0.87}$  & $-0.61^{+0.58}_{-0.54}$ & $1.9^{+2.4}_{-1.9}$ & $-0.55^{+1.63}_{-1.62}$ \\

$Joint$   & $66.96_{-0.92}^{+0.91}$  & $-2.20_{-0.98}^{+0.76}$  & $-0.67^{+0.60}_{-0.51}$ & $1.7^{+2.2}_{-1.8}$ & $-0.45^{+1.29}_{-1.26}$ \\

\hline\hline
\end{tabular}
\caption{The table shows the marginalized constrained parameters and deceleration parameter with 68\% confidence level for different datasets}
\label{tab}
\end{center}
\end{table*}

\begin{figure}[h]
\centering
\includegraphics[scale=0.70]{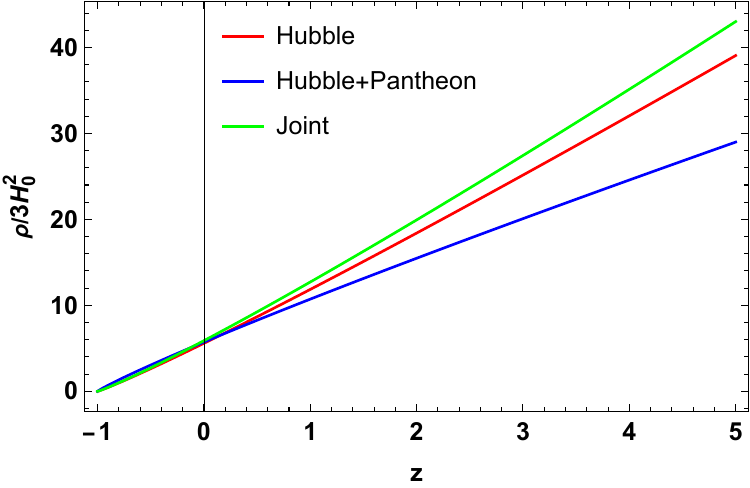}
\caption{The figure shows the evolution of density parameter as a function of redshift $z$ for the unknown parameters constrained from the Hubble, Pantheon, and Hubble+Pantheon datasets.}
\label{F_rho}
\end{figure}	

Fig. \ref{F_rho} clearly illustrates the positive behavior of the density parameter of the cosmic fluid as determined by the constrained values of the model parameters obtained from the analysis of $Hubble$, $Pantheon$, and the joint $Hubble+Pantheon$ dataset. Moreover, the density parameter is found to approach zero in the far future, indicating a promising evolution of the Universe in accordance with the constraints provided by these observational datasets.

\subsection{Deceleration parameter}
The deceleration parameter, which characterizes the rate of expansion and acceleration or deceleration of the Universe, can be expressed as
\begin{equation}
    q=-1-\frac{\dot{H}}{H^2}.
\end{equation}

Again, the deceleration parameter can be described in terms of the cosmological redshift as
\begin{equation}
    q(z) = -1+(1 + z)\frac{1}{H(z)}\frac{dH(z)}{dz}.
\end{equation}

For the current scenario, this is expressed as follows:
\begin{equation}
    q(z)=-1+\chi=-1+\left(\frac{3 (\alpha +\beta +2)}{\beta +2}+\frac{2 \alpha  \beta }{\beta +2}+\frac{(2 \beta +3) M^2}{12 (\beta +2)}\right).
    \label{DP}
\end{equation}

The deceleration parameter $q$ serves as an important indicator of the expansion dynamics of the Universe. When the deceleration parameter is greater than 0 ($q > 0$), it signifies that the Universe is experiencing a decelerated phase of expansion. This implies that the gravitational pull of matter and energy within the Universe is counteracting the expansion, causing it to slow down over time. On the other hand, if the deceleration parameter is less than 0 ($q < 0$), it corresponds to an accelerated phase of expansion. In this scenario, the repulsive effects of DE or other exotic components dominate over the gravitational pull, leading to an accelerating expansion of the Universe. The specific form of the deceleration parameter, obtained from Eq. (\ref{DP}), depends on the model parameters $\alpha$, $\beta$, and $M$. In the previous subsection, the values of these parameters were determined from different observational datasets. The corresponding value of the deceleration parameter $q$ is presented in Tab. \ref{tab}. 

\section{Discussions and Conclusions}
\label{sec5}

In this study, we have explored late-time cosmology using a well-motivated Weyl-type $f(Q,T)$ gravity model with the functional form $f(Q,T)=\alpha Q+\frac{\beta }{6\kappa ^2}T$, as proposed in \cite{Weyl1}. This model incorporates the covariant divergence of the metric tensor, which is expressed as the product of the metric and the Weyl vector $w_\mu$. The scalar of the non-metricity, represented by $Q$, is directly related to the square of the Weyl vector as $Q = -6w^2$. Therefore, the geometric properties of the theory are entirely determined by the Weyl vector and the metric tensor. The unknown parameters $\alpha$ and $\beta$ play a significant role in characterizing the behavior of the gravitational field and its interaction with the matter content of the Universe. Our analysis specifically focused on the case of dust and derived exact solutions for the field equations. As a result, we have found that the Hubble parameter $H(z)$ exhibits a power-law behavior with respect to redshift $z$.

In Sec. \ref{sec4}, we analyzed various data samples and obtained constraint values for the unknown parameters $H_0$ (Hubble constant), $\alpha$, $\beta$, and $M$ (Weyl field mass). In addition, two-dimensional likelihood contours were generated with $1-\sigma$ and $2-\sigma$ errors, representing 68\% and 95\% confidence levels for the $Hubble$, $Pantheon$, and $Hubble+Pantheon$ datasets. These contours are presented in Fig. \ref{Combine}. For the $Hubble$ dataset with 31 data points, we obtained the values $H_0=67.25_{-0.98}^{+0.97}$, $\alpha=-2.26^{+0.90}_{-1.3}$, $\beta=-0.62^{+0.59}_{-0.56}$ and $M=2.0^{+2.6}_{-2.1}$. Moving on to the $Pantheon$ dataset with 1048 sample points, we obtained $H_0=67.1_{-1.0}^{+1.1}$, $\alpha=-2.29_{-1.2}^{+0.87}$, $\beta=-0.61^{+0.58}_{-0.54}$ and $M=1.9^{+2.4}_{-1.9}$. Finally, when we combined the datasets, we found $H_0=66.96_{-0.92}^{+0.91}$, $\alpha=-2.20_{-0.98}^{+0.76}$, $\beta=-0.67^{+0.60}_{-0.51}$ and $M=1.7^{+2.2}_{-1.8}$. We compared our Weyl-type $f(Q,T)$ model with the $\Lambda$CDM model by examining the evolution of the Hubble parameter $H(z)$ and the distance modulus $\mu(z)$. We used the constraint values of the unknown parameters $H_0$, $\alpha$, $\beta$, and $M$ obtained from the $Hubble$ and $Pantheon$ datasets. The results are depicted in Figs. \ref{Hubble} and \ref{Mu}. It is evident that our Weyl-type $f(Q,T)$ model fits well with the observational results. In addition, our model closely resembles the evolution of the $\Lambda$CDM model. Therefore, the obtained value of the Hubble parameter $H_0$ in our study is consistent with the recent findings of Aubourg et al. \cite{Aubourg}, who used a combination of BAO measurements, CMB data, and a reanalysis of SNe Ia data to constrain cosmological parameters and test DE models. Further, our measurements of $H_0$ are in agreement with the results from the Planck-2018 study on the $\Lambda$CDM model \cite{Planck}. Furthermore, various DE models have also employed similar methods to obtain comparable values for the Hubble parameter $H_0$ \cite{Chen1, Chen2}.

In addition to the comparison with previous studies and the agreement with the $H_0$ value, we have also explored the behavior of the density parameter and the deceleration parameter for the constrained values of the model parameters. The energy density, as shown in Fig. \ref{F_rho}, exhibits a positive behavior, which is in line with our expectations. This behavior indicates that the energy content of the Universe remains positive and is consistent with the expansion of the Universe. The current value of the deceleration parameter is determined to be $q_=-0.46^{+2.14}_{-2.19}$ for the $Hubble$ dataset, $q_0=-0.55^{+1.63}_{-1.62}$ for the $Pantheon$ dataset, and $q_0=-0.45^{+1.29}_{-1.26}$ for the joint $Hubble+Pantheon$ dataset, and thus in line with the current understanding of the Universe undergoing accelerated expansion. 
Several studies have investigated the present value of the deceleration parameter using different approaches and datasets. Rani et al. \cite{Rani} performed a joint analysis of age of galaxies, strong gravitational lensing, and SNe Ia data, finding a value of $q_0=-0.52\pm 0.12$. Xu et al. \cite{XuDP} utilized 307 SNe Ia, along with BAO and $H(z)$ data, and obtained $q_0= -0.715\pm 0.045$. On the other hand, Holanda et al. \cite{Holanda}, used galaxy clusters with elliptical morphology, employing the Sunyaev-Zeldovich effect (SZE) and X-ray observations, and found $q_0= -0.85_{-1.25}^{+1.35}$. These studies showcase the variety of methodologies and datasets employed to constrain the deceleration parameter, emphasizing the agreement with our own study.

In conclusion, our analysis suggests that the Weyl-type $f(Q,T)$ theory provides a viable alternative to explain the late-time acceleration of the Universe without the need to introduce DE. The obtained constraint values for the model parameters support the viability of this theory in describing the observed cosmic acceleration. This opens up new avenues for further investigations and challenges the conventional understanding of the nature of DE. Finally, there is potential for exploring alternative functional forms of Weyl-type $f(Q,T)$ using similar approaches, which is left for investigation in our future research endeavors.

\textbf{Data availability} There are no new data associated with this
article.


\begin{thebibliography}{99}

\bibitem{Riess/1998} A.G. Riess et al., \textit{Astron. J.}, \textbf{116} 1009 (1998).

\bibitem{Riess/2004} A.G. Riess et al., \textit{Astophys. J.}, \textbf{607} 665-687 (2004).

\bibitem{Perlmutter/1999} S. Perlmutter et al.,  \textit{Astrophys. J.}, \textbf{517} 377 (1999).

\bibitem{Komatsu/2011} E. Komatsu et al., \textit{Astrophys. J. Suppl.}, \textbf{192}, 18 (2011).

\bibitem{Huang/2006} Z.Y. Huang et al., \textit{JCAP}, \textbf{0605}, 013 (2006).

\bibitem{Koivisto/2006} T. Koivisto, D.F. Mota, \textit{Phys. Rev. D}, \textbf{73}, 083502 (2006).

\bibitem{Daniel/2008} S.F. Daniel, \textit{Phys. Rev. D}, \textbf{77} 103513 (2008).

\bibitem{Bambar} K. Bamba, S. Capozziello, S. Nojiri, S. D. Odintsov, \textit{Astrophys.Space Sci.} \textbf{342}, 155-228 (2012). 

\bibitem{Clifton} T. Clifton et al., \textit{Phys. Rep.}, \textbf{513} 1 (2012).

\bibitem{Peebles/2003} P. J. E. Peebles and B. Ratra, \textit{Rev. Mod. Phys.}, \textbf{75}, 559 (2003).

\bibitem{Sahni/2000} V. Sahni, A. Starobinsky, \textit{Int. J. Mod. Phys. D}, \textbf{9}, 373 (2000).

\bibitem{CdM1} Sh. Nojiri, S. D. Odintsov, \textit{J.Phys.A} \textbf{40}, 6725-6732 (2007).

\bibitem{CdM2} P.K.S. Dunsby, E. Elizalde, R. Goswami, S.Odintsov, D. Saez Gomez, \textit{Phys.Rev.D} \textbf{ 82}, 023519 (2010).

\bibitem{Nojiri2011} S. Nojiri and S. D. Odintsov, \textit{Phys. Rep.}, \textbf{505}, 59-144 (2011).

\bibitem{Carroll2004} S. M. Carroll et al., \textit{Phys. Rev. D}, \textbf{70}, 043528 (2004).

\bibitem{Harko2011} T. Harko, F. S. N. Lobo, S. Nojiri, and S. D. Odintsov, \textit{Phys. Rev. D}, \textbf{84}, 024020 (2011).

\bibitem{Koussour1} M. Koussour and M. Bennai, \textit{Int. J. Geom. Methods Mod.}, \textbf{19}, 2250038 (2022).

\bibitem{B2} M. F. Shamir, \textit{Eur. Phys. J. C} \textbf{75}, 8 (2015).

\bibitem{Felice2009} A. De Felice and S. Tsujikawa, \textit{Phys. Lett. B}, \textbf{675}, 1-8 (2009).

\bibitem{Koussour2} M. Koussour et al., \textit{Nucl. Phys. B.}, \textbf{978}, 115738 (2022).

\bibitem{Hayashi} K. Hayashi and T. Shirafuji, \textit{Phys. Rev. D}, \textbf{19} 3524 (1979).

\bibitem{bengocheu09} G. R. Bengochea, R. Ferraro, \textit{Phys. Rev. D}, \textbf{79} 124019 (2009).

\bibitem{linder10} E. V. Linder, \textit{Phys. Rev. D}, \textbf{81} 127301 (2010).

\bibitem{Koussour3} M. Koussour and M. Bennai, \textit{Class. Quantum Gravity}, \textbf{39}, 105001 (2022).

\bibitem{Benetti} M. Benetti, S. Capozziello, and G. Lambiase, \textit{Mon. Not. R. Astron. Soc.}, \textbf{500}, 1795 (2021).

\bibitem{fQ1} J. B. Jimenez, L. Heisenberg, T. Koivisto, \textit{Phys. Rev. D}%
, \textbf{98}, 044048 (2018).

\bibitem{fQ2} J. B. Jimenez et al., \textit{Phys. Rev. D}, \textbf{101}, 103507 (2020).

\bibitem{fQ4} S. Capozziello and M. Shokri, \textit{	Phys. Dark Univ.}, \textbf{37}, 101113 (2022).

\bibitem{fQ5} F. Bajardi and S. Capozziello, \textit{Eur. Phys. J. C}, \textbf{83}, 531 (2023).

\bibitem{fQ6} S. Capozziello and R. D'Agostino, \textit{Phys. Lett. B}, \textbf{832}, 137229 (2022).

\bibitem{fQ7} M. Koussour et al., \textit{Fortschr. Phys.}, \textbf{71}, 2200172 (2023).

\bibitem{fQ8} A. Mussatayeva et al., \textit{Phys. Dark Univ.}, \textbf{42}, 101276 (2023).

\bibitem{fQ3} H. Weyl, \textit{Sitzungsber. Preuss. Akad.Wiss.}, \textbf{465}, 1 (1918).

\bibitem{fQT1} Y. Xu et al., \textit{Eur. Phys. J. C} \textbf{79}, 1-19 (2019).

\bibitem{QT2} S. A. Narawade, M. Koussour, and B. Mishra, \textit{Nucl. Phys. B.} \textbf{992}, 116233 (2023).

\bibitem{QT3} M. Koussour et al., \textit{Int. J. Mod. Phys. D} \textbf{31}, 2250115 (2022).

\bibitem{QT4} A. Nájera and A. Fajardo, \textit{J. Cosmol. Astropart. Phys.} \textbf{2022}, 020 (2022).

\bibitem{QT5} M. Shiravand, S. Fakhry, and M. Farhoudi, \textit{Phys. Dark Universe} \textbf{37}, 101106 (2022).

\bibitem{QT6} K. E. Bourakadi et al., \textit{Phys. Dark Universe} \textbf{41}, 101246 (2023).

\bibitem{QT7} S. Arora et al., \textit{Phys. Dark Universe} \textbf{30}, 100664 (2020).

\bibitem{QT8} T. H. Loo, M. Koussour, and A. De, \textit{Ann. Phys.} \textbf{454}, 169333 (2023).

\bibitem{QT9} M. Tayde et al., \textit{Chin. Phys. C} \textbf{46}, 115101 (2022).

\bibitem{Weyl1} Y. Xu et al., \textit{Eur. Phys. J}. C \textbf{80}, 449 (2020).

\bibitem{Weyl2} Jin-Z. Yang et al., \textit{Eur. Phys. J. C} \textbf{81},
111 (2021).

\bibitem{Weyl3} M. Koussour, \textit{Chin. J. Phys.} \textbf{83},
454-466 (2023).

\bibitem{Mackey/2013} D. F. Mackey et al., \textit{Publ. Astron. Soc. Pac.} 
\textbf{125}, 306 (2013).

\bibitem{Yu/2018} Yu, B. Ratra, F-Yin Wang, Astrophys. J., \textbf{856}, 3
(2018).

\bibitem{Moresco/2015} M. Moresco, Month. Not. R. Astron. Soc., \textbf{450}, L16-L20 (2015).

\bibitem{Sharov/2018} G.S. Sharov, V.O. Vasilie, Mathematical Modelling and Geometry \textbf{6}, 1 (2018).

\bibitem{Scolnic/2018} D.M. Scolnic et al., Astrophys. J, \textbf{859}, 101
(2018).

\bibitem{Chang/2019} Z. Chang et al., Chin. Phys. C, \textbf{43}, 125102
(2019).

\bibitem{Planck} Planck Collaboration, Astron. Astrophys., \textbf{641},
A6 (2020).

\bibitem{Aubourg} E. Aubourg et al.,\textit{Phys. Rev. D} 
\textbf{92}, 123516 (2015).

\bibitem{Chen1} G. Chen and B. Ratra ,\textit{PASP} 
\textbf{123}, 1127 (2011).

\bibitem{Chen2} G. Chen, S. Kumar and B. Ratra,\textit{Astrophys. J.}  \textbf{835}, 86 (2017).

\bibitem{Rani} N. Rani et al. ,\textit{J. Cosmol. Astropart. Phys.}  \textbf{12}, 045 (2015).

\bibitem{XuDP} L. Xu, W. Li, and J. Lu. ,\textit{J. Cosmol. Astropart. Phys.}  \textbf{2009}, 031 (2009).

\bibitem{Holanda} R. F. L. Holanda, J. S. Alcaniz and J. C. Carvalho,\textit{J. Cosmol. Astropart. Phys.}  \textbf{2013}, 033 (2013).

\end{thebibliography}
\end{document}